\documentclass[%
aip,jcp,amsmath,amssymb,
preprint,%
citeautoscript,
]{revtex4-2}

\usepackage[utf8]{inputenc}

\usepackage{graphicx}
\usepackage{caption}
\usepackage{subcaption}
\usepackage{siunitx}
\usepackage{booktabs}
\usepackage{xcolor}
\usepackage{soul}
\usepackage{wrapfig}
\usepackage{siunitx}

\begin{document}


\title{Generative deep learning as a tool for inverse design of high-entropy refractory alloys}

\author{Arindam Debnath}

\affiliation{Department of Materials Science and Engineering, Pennsylvania State University, University Park, PA 16802}

\author{Adam M. Krajewski}
\affiliation{Department of Materials Science and Engineering, Pennsylvania State University, University Park, PA 16802}

\author{Hui Sun}
\affiliation{Department of Materials Science and Engineering, Pennsylvania State University, University Park, PA 16802}

\author{Shuang Lin}
\affiliation{Department of Materials Science and Engineering, Pennsylvania State University, University Park, PA 16802}

\author{Marcia Ahn}
\affiliation{Department of Materials Science and Engineering, Pennsylvania State University, University Park, PA 16802}

\author{Wenjie Li}
\affiliation{Department of Materials Science and Engineering, Pennsylvania State University, University Park, PA 16802}

\author{Shashank Priya}
\affiliation{Department of Materials Science and Engineering, Pennsylvania State University, University Park, PA 16802}

\author{Jogender Singh}
\affiliation{Applied Research Laboratory, Pennsylvania State University, University Park, PA 16802}

\author{Shunli Shang}
\affiliation{Department of Materials Science and Engineering, Pennsylvania State University, University Park, PA 16802}

\author{Allison M. Beese}
\affiliation{Department of Materials Science and Engineering, Pennsylvania State University, University Park, PA 16802}

\author{Zi-Kui Liu}
\affiliation{Department of Materials Science and Engineering, Pennsylvania State University, University Park, PA 16802}

\author{Wesley F. Reinhart}
\email[email:]{reinhart@psu.edu}
\affiliation{Department of Materials Science and Engineering, Pennsylvania State University, University Park, PA 16802}
\affiliation{Institute for Computational and Data Sciences, Pennsylvania State University, University Park, PA 16802}

\date{\today}
\begin{abstract}
Generative deep learning is powering a wave of new innovations in materials design.
In this article, we discuss the basic operating principles of these methods and their advantages over rational design through the lens of a case study on refractory high-entropy alloys for ultra-high-temperature applications.
We present our computational infrastructure and workflow for the inverse design of new alloys powered by these methods.
Our preliminary results show that generative models can learn complex relationships in order to generate novelty on demand, making them a valuable tool for materials informatics.
\end{abstract}


\maketitle

\section{Introduction}

More than half of the National Academy of Engineering’s 14 Grand Challenges for the 21st Century \cite{NAE2017} involve the design, manufacture, and maintenance of advanced materials, whose functions and properties will be derived from their internal structure.
The relationship between structure and function is challenging to understand and even harder to predict because it is nonlinear, high-dimensional, and results from physical phenomena at many scales.
Traditional materials design has relied on human intuition to interpret patterns in known structure-property relationships and infer new materials with similar or improved properties.
However, as materials chemistry and processing become more complex, these strategies become increasingly challenging, and progress is stymied by an overwhelming design space.

Fortunately, new mathematical frameworks and powerful hardware to implement them have been developed to handle such difficult scientific problems.
Deep neural networks (DNNs) can learn incredibly complex nonlinear functions on text, images, and graphs \cite{goodfellow2016deep}.
DNNs extract the so-called latent features from high-dimensional input data to make meaningful transformations on them.
For example, a DNN trained to generate realistic images of human faces may learn latent features describing hair color and facial expression \cite{Tian2018}.
Thus, the model can not only be asked to generate an image with precisely the desired characteristics, expression, and lighting, but it can also ``explain'' the image to some degree.
The idea of latent spaces is not unique to machine learning; the highly influential Materials Genome Initiative (MGI) has made use of a very similar concept to revolutionize the way researchers approach rational materials design.
In the language of MGI, a material genome is a quantitative description of the underlying features of a material which govern its properties.
Likewise, the latent space of the model is a learned representation that captures the dominant modes of the variation in the observed data which lead to the variation in the properties.


While predictions about material properties can be made using traditional computational methods, an exciting and powerful new capability afforded by DNNs is the ability to approximate inverse functions.
By training a DNN to invert random noise from a prescribed distribution to approximate an observed distribution, a generative model is produced.
Once trained, such a model can draw novel samples from random noise, creating entirely new observations that approximately match the general rules from the training data without exactly matching them.
Generative models have recently been applied to a variety of materials including organics and inorganics \cite{sanchez2018inverse, bhowmik2019perspective}.
For instance, they were recently used to design composite materials with toughness exceeding 20\% of what has been achieved through other optimization methods (e.g., topology optimization) \cite{chen2020generative}.
Similar approaches have been demonstrated for optical meta-materials \cite{yeung2020global} and bulk\cite{dan2020generative} and thin-film\cite{dong2020inverse} inorganic materials.
Aside from the design of new materials, generative models are also becoming a popular method for reconstructing high-resolution images from partial or noisy microscopy data  \cite{iyer2019conditional}.

Here we will consider a case study on a particular class of materials, high entropy refractory alloys \cite{senkov2018development}.
We discuss the challenges in using traditional design schemes, even those accelerated by recent machine learning approaches, and how generative deep learning can provide solutions.
We describe the data ecosystem that enables our approach and provide preliminary results from the generative models trained on those data.
Finally, we conclude with some brief remarks on the future challenges in applying these techniques to materials design.

\section{Design of high-entropy refractory alloys}
Ni-based superalloys have been a popular material system for high temperature applications like turbines due to their exceptional properties at elevated temperatures.
However, the current generation of Ni-based components are operating at close to their melting point (1100°C) \cite{senkov2018development}, and additional thermal management strategies such as internal cooling channels and conventional thermal barrier coatings have also been pushed to their limits.
The ability to operate at even higher temperatures will lead to an increase in the efficiency of these systems and lead to a reduction in carbon emission and an increase in fuel and energy saving.
Therefore, there has been an increase in the demand for new materials that display superior mechanical properties at temperatures as high as 1600°C.

Refractory alloys are promising candidates as they exhibit desirable properties at elevated temperatures.
However, traditional refractory alloys also exhibit low ductility at room temperature and are prone to oxidation \cite{philips2020new}.
A variety of processing techniques have been employed in attempts to address these drawbacks \cite{philips2020new, melia2020high}.
A different route is to produce High-Entropy Alloys (HEAs) from the refractory elements \cite{chen2018review, senkov2018development}.
However, a very limited number of HEAs that surpass the performance of Ni-based superalloys have been discovered so far. Designing new HEAs that meet these requirements using the conventional trial-and-error approach is therefore a challenging task that not only requires domain knowledge but also depends on fortuitous discovery.

\subsection{Data-driven rational design}

Computational tools for prediction and evaluation of stable phases based on thermodynamics using the CALculation of PHAse Diagram(CALPHAD) approach and first-principles in terms of the Density Functional Theory (DFT) have matured in the last decade and continue to contribute to an increasingly rich ecosystem of data \cite{Liu2018}.
Well populated databases of alloy phase stability can enable rational design through expert intuition or more sophisticated numerical techniques \cite{Li2018, Wu2018}.
The quantity and span of these computational methods has the potential to greatly reduce the barrier to the rational, forward design of improved materials.
These datasets can guide experimental synthesis to the most promising candidates, leading to substantially better materials from only a handful of experiments \cite{Wen2019}.
However, there is more work to be done on making these data accessible to the general scientific community through software for data mining and predictive modeling.

Based on these plentiful datasets, machine learning approaches such as deep learning can be deployed to rapidly predict the properties of hypothetical compounds \cite{Krajewski2020, Tawfik2020, Chibani2020, Goodall2020, Schleder2019, Schmidt2019}.
Targeted alloy design can be achieved by surrogate models for specific material properties \cite{Dai2020, Kim2019, Qu2019}.
While such methods have been successfully employed, for instance, to synthesize new Co-based alloys \cite{Yu2019, Ruan2020}, they still have to rely on a human designer to properly utilize the forward-mode surrogate models. This human can help introduce some valuable expert-knowledge into the workflow, but at the same time, slows down the overall process and can introduce unintended bias. 

High-entropy alloy (HEA) design specifically has benefited from data-driven modeling in recent years.
In this case, data-driven design refers to optimization or improvement of material properties such as stability, hardness, or manufacturability with the help of surrogate models \cite{Jha2018, Nomoto2020}.
The most straightforward of these approaches take advantage of the availability of historical experimental and computational data, while more sophisticated implementations include the design of experiments and simulation in the loop.
For instance, a variety of data-driven methods have been used to predict the stable phases of HEAs in recent years \cite{Huang2019, Li2019, Qu2019, Kaufmann2020} with particular attention on single-phase HEAs.
Unfortunately, even with the success of these forward models, the conventional combinatorial approach to candidate selection leaves a design space discouragingly large to probe in the case of equiatomic HEAs \cite{Kaufmann2020}, or physically impossible to investigate completely in the case of non-equiatomic HEAs.

\subsection{Generative modeling}

We aim to build on recent success in end-to-end DNN architectures used in other material design contexts which rely on implicit feature learning \cite{Flam-Shepherd2020, Kim2020}.
A core advantage of these models is the ability to learn meaningful representations of complex design spaces.
The learned spaces are low-dimensional and smooth by construction (i.e., using a normal random vector), whereas the original design spaces may be jagged and discontinuous in many dimensions.

The most popular variety of these models is the Generative Adversarial Network (GAN)\cite{goodfellow2016nips}.
A GAN model consists of two DNNs: a generator that learns a mapping between a random normal latent space and the target distribution (effectively generating new data), and a critic that learns to distinguish between the real observations and generated data from its adversary.
The term ''adversarial'' refers to the training procedure in which the two networks compete with each other, the generator trying to produce increasingly realistic examples and the discriminator trying to catch the generator in the act.
This scheme allows the generator to learn very high quality representations without much training data.

\subsection{Towards inverse design}

\begin{figure}
    \centering
    \vspace{12pt}
    \includegraphics[width=\linewidth]{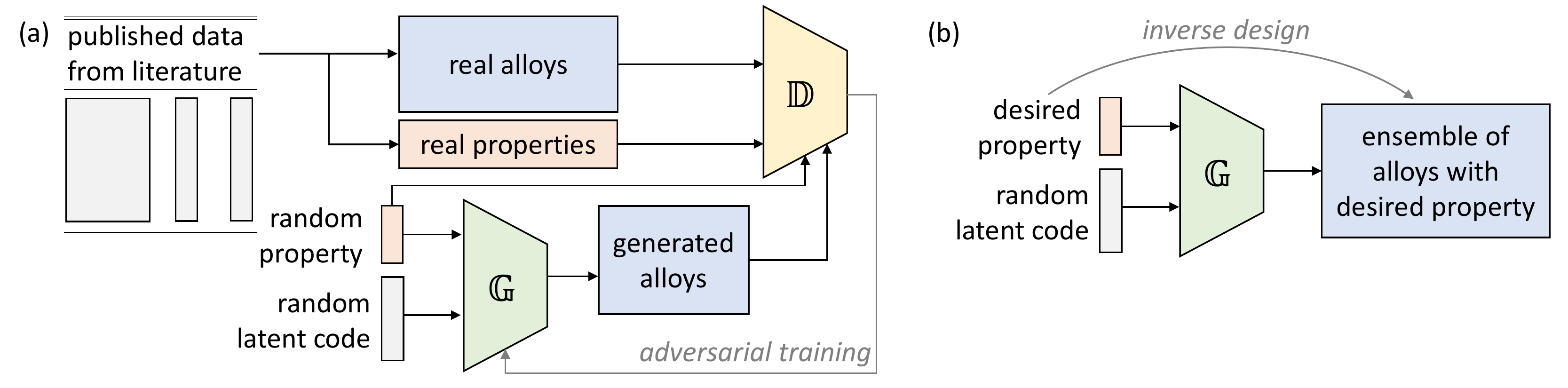}
    \caption{Schematic illustration of generative modeling for inverse design of materials using a conditional GAN.
    (a) Adversarial training procedure in which the Generator and Discriminator compete for superior performance.
    (b) Inverse design using the trained Generator.}
    \label{fig:cgan-schematic}
\end{figure}

In vanilla GAN, there is no way to control the output produced by the generator, meaning that many samples must be drawn before a suitable candidate is found.
This can be controlled in the conditional GAN (cGAN) architecture, in which the generator is provided with an additional conditioning vector that enforces a mapping between the latent space and the desired figure of merit \cite{aggarwal2019regression}.
In this way, the generator learns the probability distributions of the underlying alloy properties data conditioned on the alloy composition, and therefore, samples drawn from the multi-dimensional distribution will represent viable compositions with predictable properties.
The scheme is illustrated in Figure \ref{fig:cgan-schematic}.

The cGAN approach has been demonstrated on the design of Al alloys with validation by computational methods \cite{Nguyen2018}.
In that case, the use of conditional density estimation in the inverse problem enables extremely efficient exploration of a high-dimensional design space resulting in the design of dozens of new stable alloys.
The success of these models for solving design problems relies heavily on the property of invertibility, which means that promising points in the latent space can be sent through the model in reverse to yield candidates in the original design space.
Access to an invertible latent space enables rapid candidate material generation with the ability to interpolate continuously between desirable structures, as demonstrated with Metal-organic Frameworks (MOFs) \cite{Yao2020}, rather than the more rudimentary combinatorial high-throughput screening associated with forward design methods.

There are a variety of alternative approaches which could be considered for this problem.
Without generative architectures, the design process would typically proceed in two stages.
First, supervised learning could be used to train predictive models for the properties of interest.
Using this fast surrogate model, optimization (e.g., gradient descent) could then be performed to identify an input composition to yield the desired properties.
This is generally not preferred since generative models can produce suitable compositions in a single step.

It is noted that there are other generative architectures besides GAN that are viable for this problem, such as the conditional variational autoencoder (cVAE) \cite{lim2018molecular}.
VAEs minimize a reconstruction loss to learn a suitable latent space instead of relying on adversarial training to learn the mapping from a reference distribution to the distribution of interest as GANs do.
However, VAEs have been shown to produce inferior results to GANs due to the noise injection inherent to the training procedure and the requirement of a predefined metric for reconstruction error \cite{bao2017cvae}.

Despite their advantages, it is known that cGANs are difficult to work with and require significant tuning to obtain good results.
In the training procedure, a suitable distribution for the conditioning vector must be provided to ensure that both the generator and discriminator have opportunities to explore the joint distribution.
These models can also suffer from vanishing gradient, convergence problems, and mode collapse \cite{goodfellow2016nips}.
While strategies such as Wasserstein GAN~\cite{arjovsky2017wasserstein} offer piecemeal solutions, ultimately GAN remains a convenient approximation rather than a cure-all solution to implicit data modeling \cite{li2018implicit}.

\section{Case study: inverse design of refractory HEAs}

\subsection{Data ecosystem}

\begin{figure}[ht]
    \centering
    \includegraphics[width=0.75\textwidth]{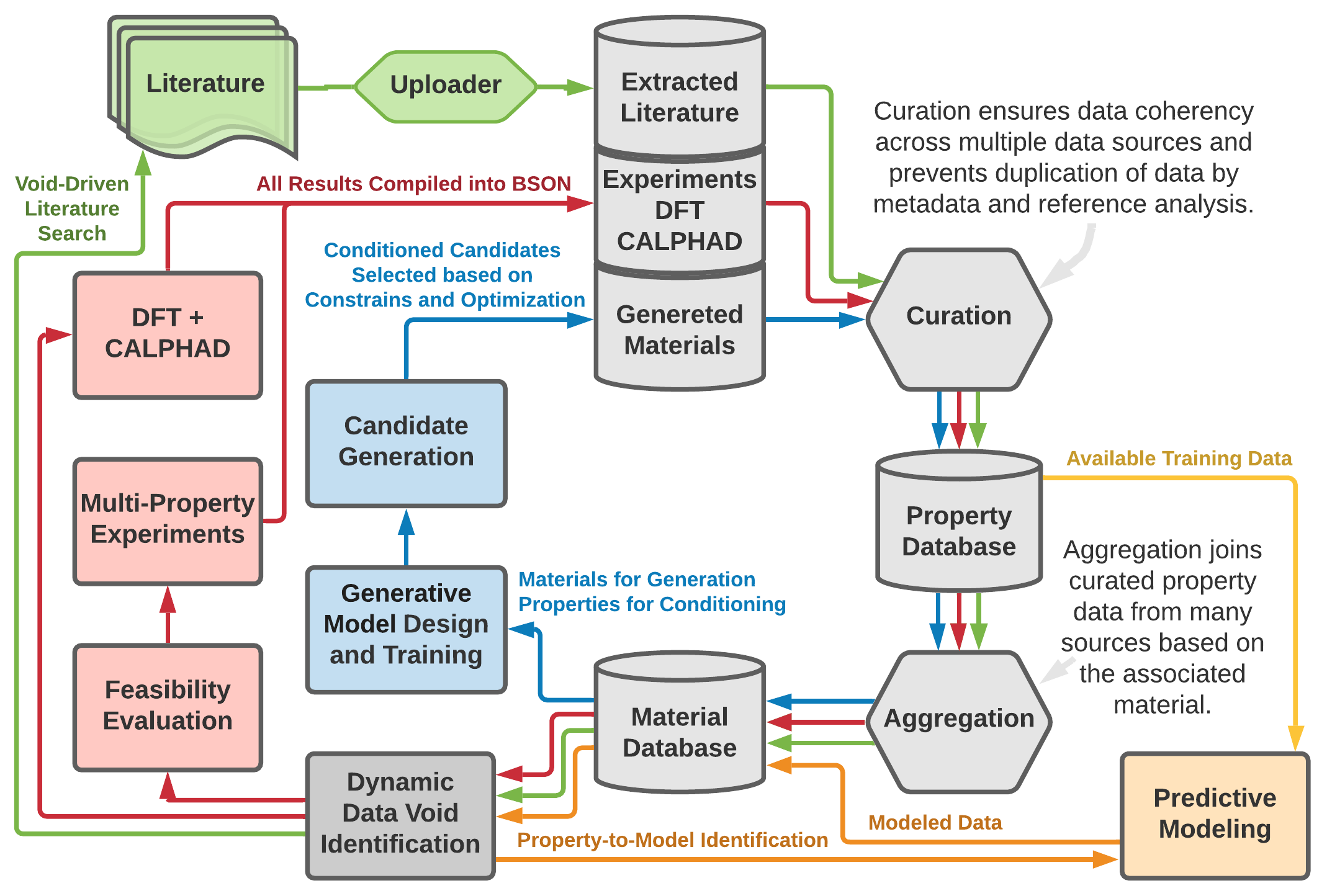}
    \caption{Data ecosystem schematic.}
    \label{fig:ecosystem}
\end{figure}

Any generative material design effort requires close integration with existing literature data and scientific techniques to validate generated samples beyond the known set. In this case study, we accomplish this by creating an advanced data ecosystem, presented in Figure \ref{fig:ecosystem}. It seamlessly merges literature, validation, and generated data by retaining their independence at the single data point level, yet ensuring a coherent JavaScript Object Notation (JSON)-like data representation and combining them at the single unique material level, as shown in the gray section of Figure \ref{fig:ecosystem}.


This arrangement, centered around automated identification of unique materials, allows an efficient and fully automated identification of voids in the current state of database knowledge. These voids can then be dealt with dynamically by the appropriate component of the ecosystem every time a change in the database is detected, e.g., whenever a new alloy is designed by a GAN. In this case study, this is accomplished by a constantly running cloud Virtual Machine (VM) server linked to the database through a high-throughput Application Programming Interface (API). Identified missing literature data is passed to Natural Language Processing (NLP)-based search algorithms and researchers, who attempt to fill it (green loop in  Figure \ref{fig:ecosystem}). 

\begin{figure}
    \centering
    \includegraphics[width=0.43\textwidth]{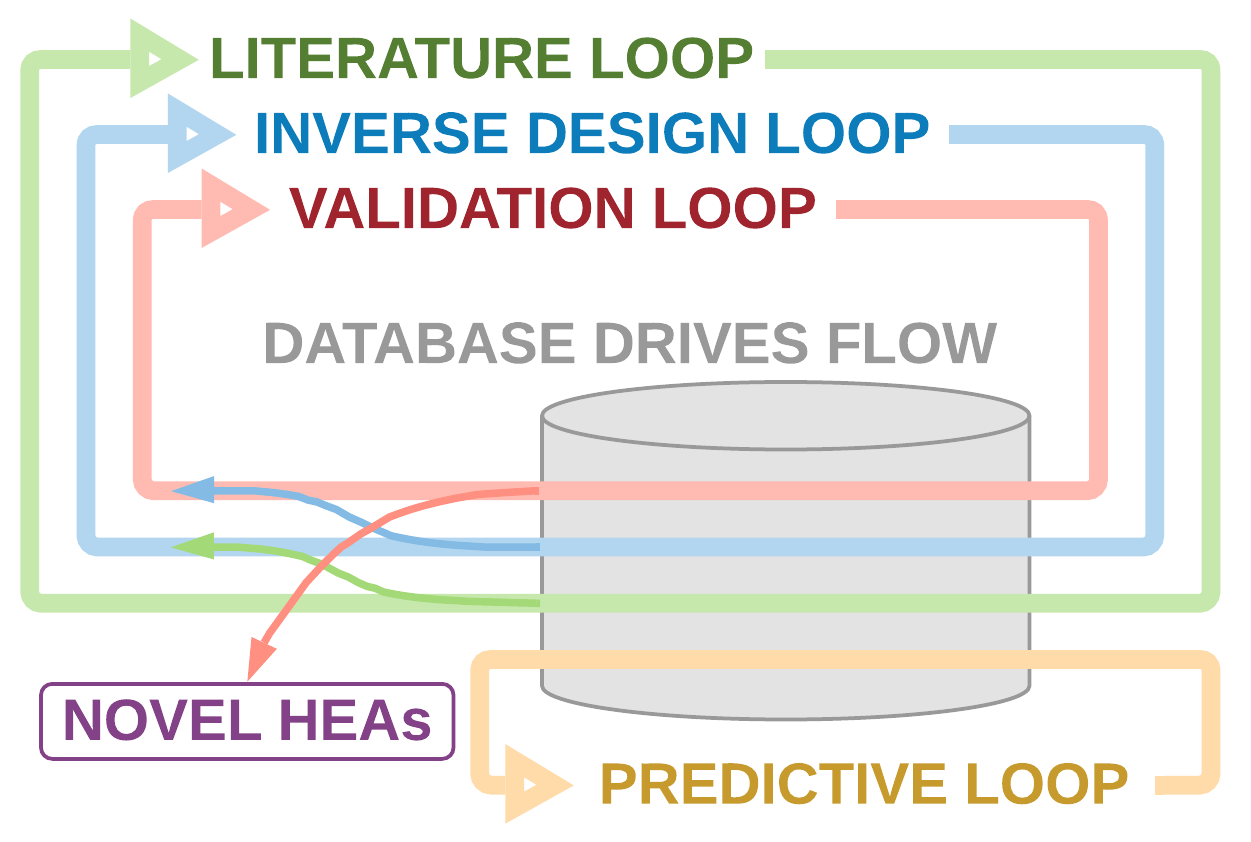}
    \caption{Four main data flow paths in the ecosystem.}
    \label{fig:dataflow}
\end{figure}

Data identified as missing a necessary validation is passed to computational techniques and researchers responsible for experiments (red loop). At the same time, predictive models attempt to rapidly fill in any void with approximations (orange loop) based on all defined empirical models from the literature and data-driven predictions based on already known data. In this case study, the structure-aware linear combination of elemental properties was found to be particularly useful. A void-free dataset of materials with various properties is then employed to create generative models, with materials used as samples and associated properties used for conditioning the model.
With trained GANs, new candidates are generated and uploaded back to the low-level dataset as novel materials in need of validation.
We describe this generation process in detail in the following sections.
This ecosystem design inherently leads to a data flow within independent yet interacting loops, shown in Figure \ref{fig:dataflow}, providing many benefits to the design process. Foremost, it allows interaction between literature, inverse design, and validation to be fully automated, making sure that at any given time, GANs are trained on all available data and validations are run on the most recent candidate selection. Once running, it eliminates any wait stages resulting in maximization of discovery rate given resources.


\subsection{Building a generative model}

Once a sufficient dataset was collected in the literature loop shown in Figure \ref{fig:dataflow}, we began to fuel the inverse design component of the data ecosystem. To demonstrate novel refractory HEAs with the desired properties, a cGAN model based on a simple feedforward NN architecture with 4 fully connected layers was trained using 529 HEA literature-derived compositions from our database \cite{ULTERA}.
The cGAN was conditioned on the shear modulus and fracture toughness values so that we can later generate new compositions which should exhibit specific values of these properties. The values of these properties were normalized to ensure that the importance of each feature is equivalently reflected on the model. The conditioning values were sampled using the probability distribution of the property values. Batches of normally distributed sixteen dimensional latent vectors and the sampled conditioning vectors were then provided as input to the generator. 
One advantage of adversarial loss of GANs over other competing methods like reconstructive loss of VAEs is the simplicity of the objective function -- here the generator receives the negative critic score as its loss, such that it maximizes the ``realism'' of the generated samples.
Because the critic is trained in tandem with the generator, there is no need to define a metric for this ``realism,'' and it is learned directly from the observed distribution. 
We used the Wasserstein GAN\cite{arjovsky2017wasserstein} loss to avoid vanishing gradients and the unrolled GAN\cite{metz2016unrolled} strategy to avoid mode collapse.
Training the model took about one hour on an NVIDIA Tesla P100 GPU.

The properties of the generated material compositions will next be verified experimentally or through other computational approaches such as ab-initio DFT-based calculations combined with CALPHAD models\cite{Liu2009}, and fed back into the data ecosystem to serve as new training dataset for the cGAN, as illustrated in Figure~\ref{fig:dataflow}. This cycle will ensure continuous generation of novel candidate alloys, with each iteration increasing the probability of arriving at the targeted properties.

\begin{figure}[h!]
\centering
\includegraphics[width=0.9\linewidth]{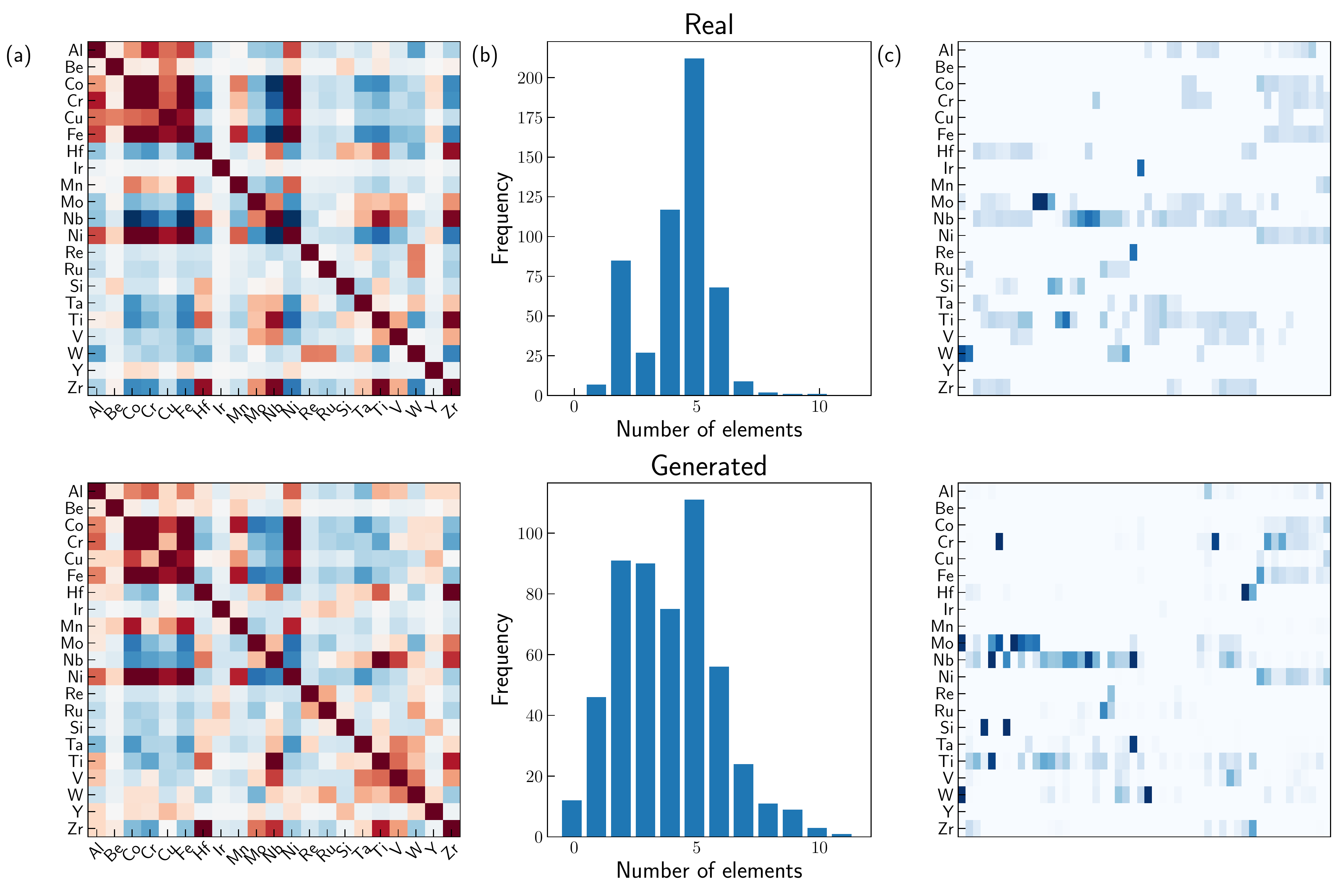}
\caption{Comparison of real (top row) and generated (bottom row) compositions.
(a) Correlation between pairs of elements. Increasing value of red indicates element pair more likely to appear in HEA composition, increasing value of blue indicates element pair less likely to appear in HEA composition
(b) Number of different elements present in each alloy.
(c) Some sample compositions. Each column represents an alloy, according to the number density of each element.The intensity of blue indicates the atomic fraction of the element in the composition.}
\label{fig:gen_real_compare}
\end{figure}

We first show that the cGAN can learn the underlying distribution of refractory HEAs; in effect, the adversarial training teaches the generator a set of design rules for what a HEA looks like.
When generating new samples, an observer should be convinced that these are legitimate alloys.
Thus, to evaluate the generator, we consider some different measures of the generated ensemble of alloy compositions in Figure~\ref{fig:gen_real_compare}.
While some minor differences can be observed, the generator appears to have largely captured the fundamental definition of a refractory HEA -- such as the correlation between different elements and the number of different constituent elements -- without requiring us to provide any guidance to the model (e.g., design rule) aside from a collection of raw data.

\begin{figure}[h!]
\centering
\includegraphics[width=0.7\linewidth]{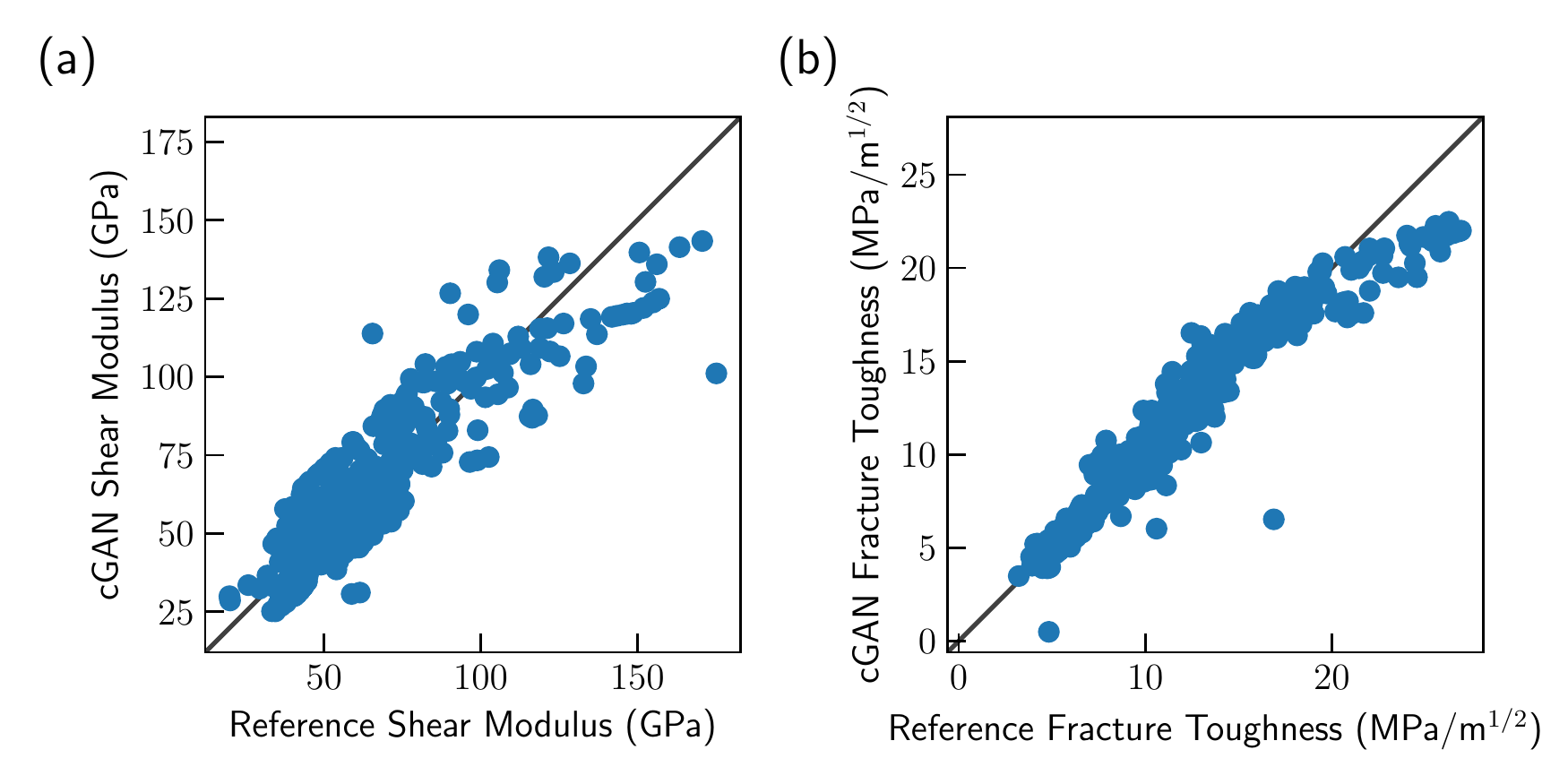}
\caption{Comparison of reference and cGAN (a) shear modulus and (b) fracture toughness values for the compositions in our database}
\label{fig:sm_ft_fit_multi}
\end{figure}

In addition to generating valid compositions, we also want to learn the joint distribution between compositions and material properties.
To evaluate this, we plot the conditioning supplied to the generator against the reference property value in Figure \ref{fig:sm_ft_fit_multi}, provided as the ground-truth.
As most reports of HEAs in the literature do not include shear modulus $G$ and fracture toughness $K_{IC}$, reference values were derived based on a linear combination (LC) of the pure elemental properties from DFT calculations \cite{Chong_2021}.
The shear modulus was approximated as a simple LC of elemental shear modulus values, while fracture toughness was obtained using Rice's model \cite{rice1992dislocation}, given by the equation
\begin{equation*}
K_{IC} = \sqrt{2\times G \times E_{USF}/(1-\nu)}   
\end{equation*}
where $E_{USF}$ is the unstable stacking fault energy, $G$ is the shear modulus for sliding along the slip plane, and $\nu$ the Poisson's ratio for the stable element reference structure.
There is good agreement in regions with more prevalent training data ($\SI{40}{\GPa} < G < \SI{100}{\GPa}$), while peripheral regions with fewer observations ( $G > \SI{100}{\GPa}$) show a weaker fit.
Overall, both the shear modulus and fracture toughness values are well captured by the cGAN model over a majority of the data domain.

\subsection{Inverse design}
\begin{figure}[h!]
\centering
\includegraphics[width=\linewidth]{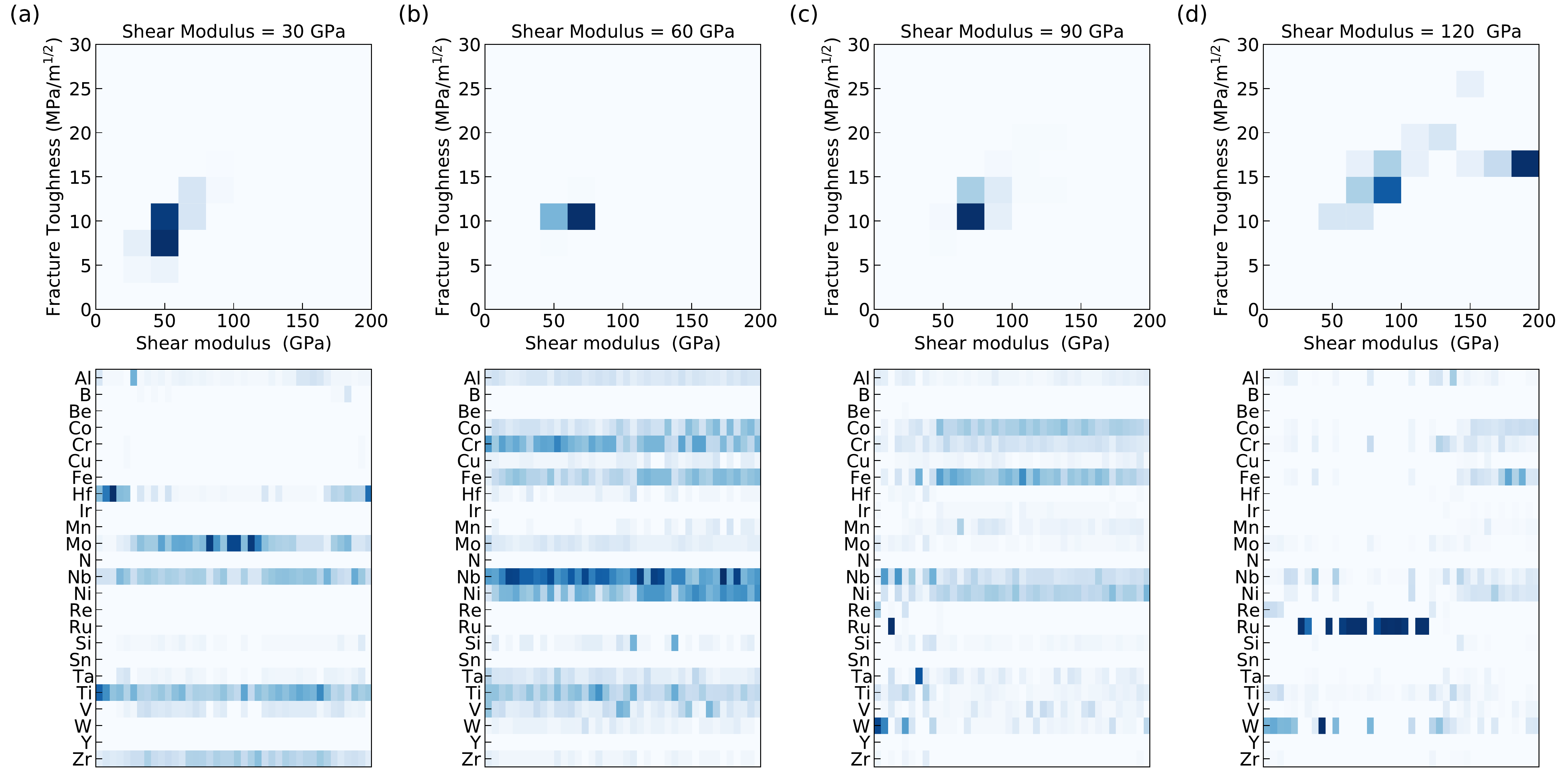}
\caption{Histograms of shear modulus and fracture toughness (top) and sample compositions (bottom) generated by fixing the shear modulus values at (a) 30 GPa, (b) 60 GPa, (c) 90 GPa, and, (d) 120 GPa.Each column represents an alloy, according to the number density of each element. The intensity of blue indicates a greater number of compositions with the corresponding values of shear modulus and fracture toughness in the top plots and the atomic fraction of the element in the composition in the bottom plots}
\label{fig:fixed_cond_single}
\end{figure}

We next demonstrate how the trained model can be used to perform inverse design of HEA compositions with respect to the shear modulus and fracture toughness.
By supplying a conditioning vector with desired property values, the generator can be biased towards compositions that are likely to exhibit those properties.
As seen in  Figure \ref{fig:fixed_cond_single}, even though the generated compositions do not produce the exact desired value of shear modulus, they do appear to come from regions of the latent space which are better aligned with the desired outcome.
This effect can be observed from the sample compositions in Figure \ref{fig:fixed_cond_single}.
With the increasing value of shear modulus, the frequency of elements like W, Re, and Ru with high elemental shear modulus (173, 150, and 149 GPa, respectively) increase, while elements like Hf, Mo, and Zr with low elemental shear modulus (30.4, 19.7, and 32.7 GPa) decrease.
Thus, the cGAN model chooses appropriate elements to generate compositions that best approach the target properties. 

While targets (a)-(c) in Figure \ref{fig:fixed_cond_single} appear reasonably well matched, the generator struggles with (d), corresponding to a shear modulus of 120 GPa.
As shown in Figure \ref{fig:sm_ft_fit_multi}(a), there are not many compositions in our training data that exhibit approximated shear modulus in excess of 100 GPa.
As a consequence, the generator is biased against creating valid compositions that match the imposed condition.
Thus, the generator resorts to creating compositions with a broad range of shear modulus values above and below the target to compensate.

\begin{figure}[htbp]
\centering
\includegraphics[width=\linewidth]{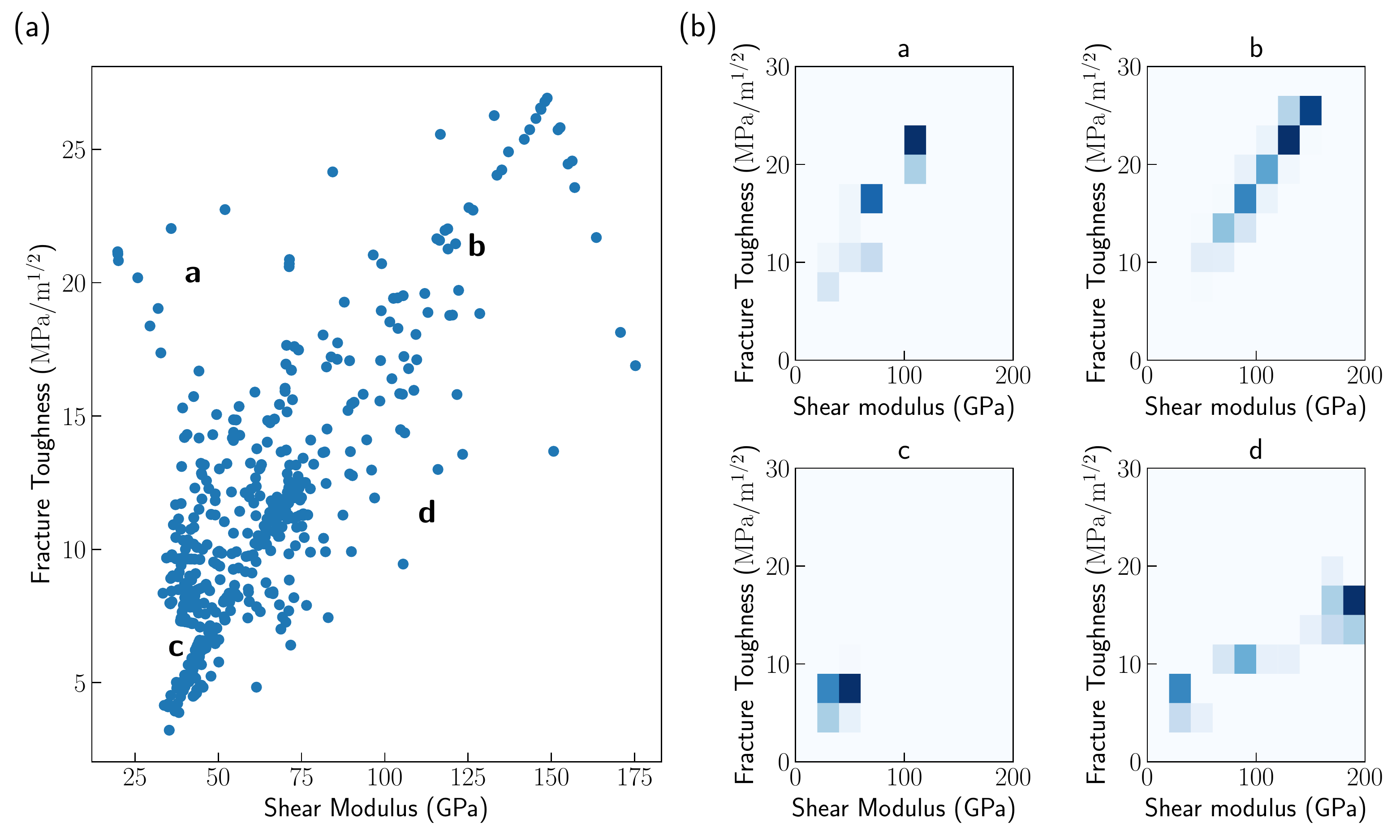}
\caption{(a) Correlation between shear modulus and fracture toughness values of the real compositions. Labels a, b, c and d represent four conditioning cases of interest.
(b) Histograms of shear modulus and fracture toughness for compositions generated using the conditions shown in panel (a). The intensity of blue in the histograms indicates a greater number of compositions with the corresponding values of shear modulus and fracture toughness}
\label{fig:correlation}
\end{figure}

Moreover, when specific values of fracture toughness are not requested from the generator, increasing the value of shear modulus naturally lead to increased fracture toughness in the generated compositions, as seen in Figure \ref{fig:fixed_cond_single}.
This is a result of the general correlation between these two properties shown in Figure \ref{fig:correlation}.
Therefore, the cGAN model implicitly learns the correlation between the shear modulus and fracture toughness values and will tend to generate compositions that have accordant values of shear modulus and fracture toughness (as shown by points b and c in Figure \ref{fig:correlation}).

\begin{figure}[h!]
\centering
\includegraphics[width=0.65\linewidth]{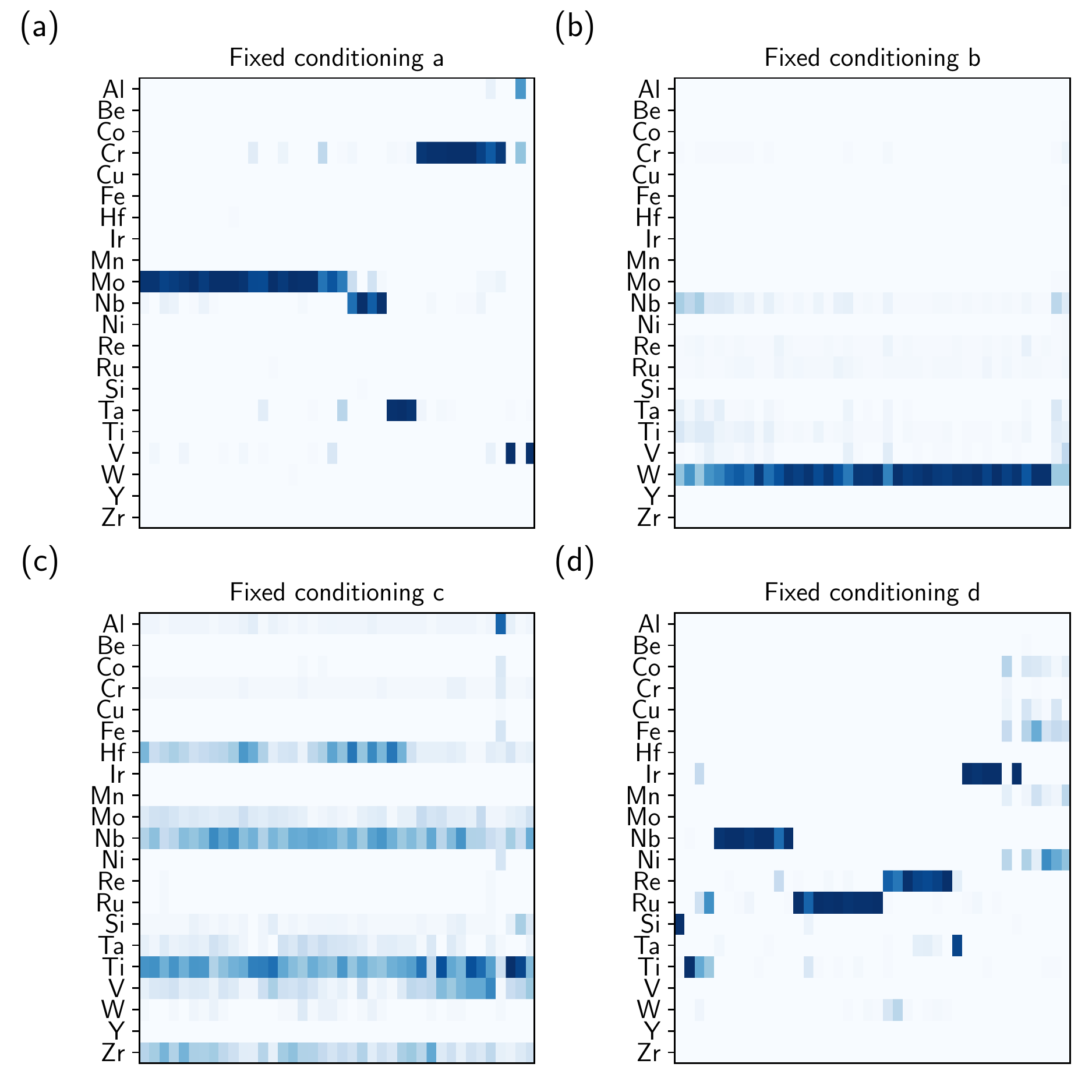}
\caption{Sample compositions generated using conditions specified in Figure~\ref{fig:correlation}. Each column represents an alloy, according to the number density of each element.The intensity of blue indicates the atomic fraction of the element in the composition.}
\label{fig:fixed_cond_bitmaps_multi}
\end{figure}

Discovering novel alloys rather than simply sampling from known compositions often requires that the cGAN model be able to generate compositions that have opposing values of these properties (e.g., high shear modulus with low fracture toughness).
To evaluate this capability, we generated an ensemble of compositions (shown in Figure \ref{fig:fixed_cond_bitmaps_multi}) with both properties specified in the conditioning vector.
This results in some interesting trends, such as more varied elemental compositions for case c and W-dominant compositions in case b.
Compositions generated using opposing conditions a and d tend to rely on a few elements like Nb and Ta in both cases while elements like Mo/Cr and Ir/Re  appear exclusively in cases a and d, respectively.
The predominance of a single element in these cases shows that the generator is relying on some particular elements with unusual properties in order to achieve these opposing objectives.

\section{Conclusions and outlook}

Generative deep learning is making an impact across a range of scientific fields, and materials informatics is no exception.
In fact, the complex relationships and high-dimensional design spaces intrinsic to materials make this a compelling domain for testing the efficacy of generative models in solving real-world problems.
Here we have shown some preliminary progress towards the inverse design of refractory high-entropy alloys using a conditional GAN.
With only a few hundred observed HEA compositions from the literature, our model was able to capture important trends in the data and reproduce realistic-looking compositions.

We demonstrated the ability of the trained model to design new alloys with targeted properties based on a learned correlation between approximated mechanical properties and the latent code used by the generator.
While it does not produce a perfect match, this conditioning strongly biases the types of compositions generated by the model.
Notably, the generator struggled when pushed to the limits of the training data domain and when the conditioning reflected rare corner cases, pointing to the gap for the need of new computational or experimental data.
This is an important obstacle to address if the model is to be used to explore new alloy compositions with exceptional properties, and points to a promising avenue of ``hybrid methods'' which use both generative deep learning models and conventional physics-based models to maximize new information gained in each iteration of computation and synthesis. 

Overall, we believe these generative models are a promising new approach to materials design which will be put to best use in conjunction with more conventional computational techniques.
In our case study of HEAs design, we employ them as an inexpensive, low fidelity approach to generate new and interesting samples which are then automatically paired with more expensive, high fidelity validation steps.
As innovation in the area of deep learning has been incredibly fast paced in recent years, in part due to large investments by industry, a key challenge to making the most of these technologies is modifying architectures developed for other problems like computer vision to work for materials design.
Ultimately this presents more opportunities than obstacles since it should allow for constantly improving models as researchers learn general strategies for model adaptation, and use them to guide other well established techniques.

\section*{Acknowledgments}

The present work is based upon work supported by the Department of Energy / Advanced Research Projects Agency – Energy (ARPA-E) under award No DE-AR0001435.

%

\end{document}